\documentclass[prl,amsmath,amssymb,twocolumn,showpacs,floatfix,groupedaddress,superscriptaddress]{revtex4}
\usepackage{graphicx}
\usepackage[english]{babel}
\usepackage{times}
\usepackage{dcolumn}
\usepackage[usenames,dvipsnames]{color}
\usepackage{units}

\begin{document}

\title{Ground state of the parallel double quantum dot system}

\author{Rok \v{Z}itko}

\affiliation{Jo\v{z}ef Stefan Institute, Jamova 39, SI-1000 Ljubljana,
Slovenia}
\affiliation{Faculty  of Mathematics and Physics, University of Ljubljana,
Jadranska 19, SI-1000 Ljubljana, Slovenia}

\author{Jernej Mravlje}

\affiliation{Centre de Physique Th{\'e}orique,
{\'E}cole Polytechnique, CNRS, 91128 Palaiseau Cedex,
France}
\affiliation{Jo\v{z}ef Stefan Institute, Jamova 39, SI-1000 Ljubljana,
Slovenia}

\author{Kristjan Haule}

\affiliation{Physics Department and Center for Materials Theory,
Rutgers University, Piscataway NJ 08854, USA}

\date{\today}

\begin{abstract}
We resolve the controversy regarding the ground state of the parallel
double quantum dot system near half filling. The numerical
renormalization group (NRG) predicts an underscreened Kondo state with
residual spin-1/2 magnetic moment, $\ln 2$ residual impurity entropy,
and unitary conductance, while the Bethe Ansatz (BA) solution predicts
a fully screened impurity, regular Fermi-liquid ground state, and zero
conductance. We calculate the impurity entropy of the system as a
function of the temperature using the hybridization-expansion
continuous-time quantum Monte Carlo technique, which is a numerically
exact stochastic method, and find excellent agreement with the NRG
results. We show that the origin of the unconventional behavior in
this model is the odd-symmetry "dark state" on the dots.
\end{abstract}

\pacs{72.10.Fk, 72.15.Qm}

\maketitle

\newcommand{\vc}[1]{{\mathbf{#1}}}
\newcommand{\braket}[2]{\langle#1|#2\rangle}
\newcommand{\expv}[1]{\langle #1 \rangle}
\newcommand{\corr}[1]{\langle\langle #1 \rangle\rangle}
\newcommand{\ket}[1]{| #1 \rangle}
\newcommand{\Tr}{\mathrm{Tr}}
\newcommand{\kor}[1]{\langle\langle #1 \rangle\rangle}
\newcommand{\degg}{^\circ}
\renewcommand{\Im}{\mathrm{Im}}
\renewcommand{\Re}{\mathrm{Re}}
\newcommand{\GG}{{\mathcal{G}}}
\newcommand{\atanh}{\mathrm{atanh}}
\newcommand{\sgn}{\mathrm{sgn}}

Quantum dot (QD) nanostructures serve as model systems for studying
fundamental many-particle effects, such as the competition between the
Kondo screening and the exchange interaction.  These effects can give
rise to ground states of the Fermi liquid or non-Fermi liquid
nature. The fingerprints of those different states as well as the
quantum-phase transitions between them have been predicted
theoretically within the generalizations of the Anderson impurity
model and observed in transport experiments \cite{goldhabergordon1998b,cronenwett1998,wiel2000,sasaki2000,jeong2001,holleitner2002,wiel2002,craig2004,chen2004,sasaki2004,potok2007,roch2008,roch2009,sasaki2009}.

To account for such a rich behavior powerful non-perturbative
theoretical tools must be used. Among these techniques, the numerical
renormalization group (NRG) \cite{wilson1975, bulla2008} is popular
because of its wide applicability, reliability, and relatively low
computational demands. The NRG results for the conductance of single
QDs in the Kondo regime have played the key role in conclusively
proving the occurrence of the Kondo effect in these systems
\cite{goldhabergordon1998a,wiel2000,costi1994}. The NRG is an
approximate method, expected to be asymptotically exact on the lowest
energy scales. In constructing the effective chain Hamiltonian, the
conductance band continuum is discretized, a decomposition into
Fourier modes is performed in each interval, and a single
representative state from each interval is retained.
This approximation is, however, well controlled
\cite{wilson1975}.  Some impurity models are integrable and can be
solved exactly using the Bethe Ansatz (BA) technique; these analytical
solutions are very valuable as reference results which serve as
benchmark for more generally applicable methods. The single-impurity
Kondo and Anderson models are both integrable
\cite{andrei1983,tsvelick1983} and an excellent agreement has been
found between the NRG and BA results \cite{andrei1983}.

\begin{figure}
\centering
\includegraphics[width=5cm]{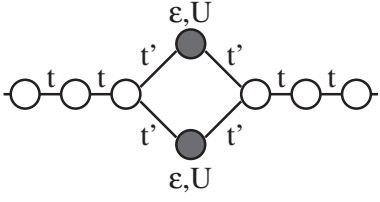}
\caption{Schematic representation of the model for a parallel double
quantum dot coupled to two semi-infinite tight-binding chains with
equal hopping constants. The Fermi level is fixed at zero energy.}
\label{dqd}
\end{figure}

One of the simplest problems that exhibits non-Fermi liquid behavior
is that of the two impurities coupled to conduction bands: the double
quantum dot (DQD).  Recently, a BA solution has been proposed for a
family of two-impurity models of DQDs
\cite{konik2007,konik2007njp,kulkarni2011,kulkarni2011n}. For two QDs
coupled in parallel between the conductance leads with equal
hybridization strengths, thus forming a symmetric ring
(Fig.~\ref{dqd}), the problem maps onto a two-impurity model with a
single effective conduction band. The BA solution predicts that near
the particle-hole symmetric point the two electrons residing on the
dots form a regular Fermi-liquid (FL) singlet state with the
conduction band electrons, that the system has zero conductance at the
particle-hole symmetric point, and that the standard Friedel sum rule
is satisfied \cite{konik2007}. All these predictions are, however, at
odds with the NRG results \cite{vojta2002,hofstetter2003,lopez2005,
  pustilnik2006, vzporedne, vzporedne2, logan2009, ding2009,
  wang2011qpt}.

On general physical grounds one expects that at some high-energy
scale the two impurity moments bind into a spin-triplet effective
state due to the presence of the ferromagnetic RKKY interaction (the
model corresponds to the $r=0$ limit of the standard two-impurity
model \cite{jayaprakash1981,jones1987,jones1988, fye1990,silva1996}),
then this spin is partially screened in the single-channel spin-$1$
Kondo effect yielding a singular Fermi liquid ground state, a residual
spin-$1/2$ magnetic moment and residual $\ln 2$ entropy
\cite{nozieres1980,koller2005,mehta2005}. The NRG results are fully
consistent with this scenario. This is also in line with the
conventional wisdom that a single screening channel can screen one
half-unit of the impurity spin
\cite{nozieres1980,andrei1983,tsvelick1983}. Furthermore, the system
reaches unitary conductance at zero temperature. The Friedel sum rule
in its standard form $\delta=(\pi/2)n_\mathrm{imp}$ is violated (there
is an additional phase shift by $\pi/2$, see
Ref.~\onlinecite{logan2009}, Sec.~IV.B). The underscreened Kondo
effect has already been experimentally observed in systems described
by two-orbital impurity models similar to the one discussed in this
work \cite{roch2008,roch2009,parks2010}.

It is disconcerting that two purportedly highly reliable methods
produce opposite results for an elementary impurity model. If the NRG
were found to be flawed, this would demand a reinvestigation of the
applicability of the method and put in question the reliability of a
large number of published theoretical results. It has been suggested
that the presumed deficiency of the NRG consists in disregarding the
higher conduction-band modes in each discretization interval and a
modified discretization has been proposed \cite{kulkarni2011}. Since
the modified discretization scheme again maps onto a single effective
screening channel, the residual moment would still not be screened,
thus this does not solve the observed discrepancy.

For this reason, in this work we resolve the controversy using an 
independent method: we perform extensive numerical simulations of the
impurity Hamiltonian using the hybridisation-expansion continuous-time
quantum Monte Carlo (CTQMC) algorithm \cite{werner2006, haule2007,
  gull2011}. This method is numerically exact, its accuracy being
limited solely by the calculation time. No approximations nor
simplifications of the model Hamiltonian need to be performed. The
price for being exact are, however, heavy computational demands. We
show that the simulated results are consistent with the NRG
calculation, thus the proposed BA solution is not correct.

The Hamiltonian is $H=H_\mathrm{band} + H_\mathrm{dots} +
H_\mathrm{hyb}$. Here $H_\mathrm{band} = \sum_{k\sigma j} \epsilon_k
c^\dag_{k\sigma j}c_{k\sigma j}$ is the conduction-band Hamiltonian
where $k$ is momentum, $\sigma=\uparrow,\downarrow$ is spin, and
$j=1,2$ indexes the leads. $H_\mathrm{dots}=\sum_{i=1}^{2} \epsilon
n_i + U n_{i\uparrow} n_{i\downarrow}$ is the quantum dot Hamiltonian.
The number operator $n_{i\sigma}$ is defined as
$n_{i\sigma}=d^\dag_{i\sigma} d_{i\sigma}$ and $n_i=\sum_\sigma
n_{i\sigma}$, $\epsilon$ is the on-site energy, and $U$ is the
electron-electron repulsion. Finally, 
\begin{equation}
H_\mathrm{hyb} = \frac{1}{\sqrt{L}} \sum_{k\sigma i} \left( V_k
d^\dag_{i\sigma} c_{k\sigma} + \text{H.c.} \right)
\end{equation}
is the coupling Hamiltonian, where $L$ is a normalization constant. 
We model the conductance leads by semi-infinite tight-binding chains
and the dots couple to the end of these chains by a hopping term,
see Fig.~\ref{dqd}. The hybridization function is defined as
$\Gamma(\omega)=2 \pi \rho(\omega) V_k^2$ with $\rho(\omega)$ the
density of states in the band and $V_k$ the coupling coefficient at
momentum $k$ that corresponds to energy $\omega$; the additional
factor 2 in the expression takes into account that there are two
leads. We have $\epsilon=2t\cos{k}$ and $V_k=t'\sin{k}$, which gives
\begin{equation}
\Gamma(\omega)=\Gamma\sqrt{1-(\omega/2t)^2},
\end{equation}
with $\Gamma=(t')^2/t$. 
In the following, we use the half-bandwidth $D=2t$ as the energy unit.
The final expression for the impurity action is
\begin{equation}
\begin{split}
S=\int_0^\beta \mathrm{d}\tau \left[
\sum_{i=(1,2),\sigma} d^\dag_{i\sigma} \left( \frac{\partial}{\partial \tau}
-\mu+\epsilon\right) d_{i\sigma} + U n_{i\uparrow} n_{i\downarrow}
\right] \\
+\frac{1}{\pi} \int_0^\beta \mathrm{d}\tau \int_0^\beta \mathrm{d}\tau'
d_{e\sigma}^\dag(\tau) \Gamma(\tau-\tau') d_{e\sigma}(\tau').
\end{split}
\end{equation}
The even/odd combination of operators is defined as
$d^\dag_{e/o,\sigma} = \left( d^\dag_{1\sigma} \pm
d^\dag_{2\sigma} \right)/\sqrt{2}$. Only the even orbital hybridizes
with the conduction band.

In this problem, there are three important energy scales
\cite{vzporedne}.  On the scale of $U$ the local moments are formed
and the dots start to behave as two spin-$1/2$ impurities.  On the
RKKY scale of $ J_\mathrm{RKKY} \sim U (\rho J_K)^2 = (64/\pi^2)
\Gamma^2/U$ the spins bind into a $S=1$ state.  Here $\rho
J_K=8\Gamma/\pi U$ quantifies the strength of the exchange coupling.
Finally, on the scale of the Kondo temperature $T_K \sim U \sqrt{\rho
  J_K} \exp\left( - 1/\rho J_K \right)$, the impurity moment is
partially screened from spin-$1$ to spin-$1/2$ in the single-channel
spin-$1$ Kondo effect \cite{nozieres1980}. 
 No other low-energy
scales are present in this problem \cite{jayaprakash1981}. The
$U/\Gamma$ ratios in experiments on quantum dots range roughly from 1
to 20, thus the three energy scales introduced above are not
necessarily well separated.

Different methods for solving impurity models are best compared by
calculating thermodynamic functions such as energy or entropy,
since all other quantities of interest can be obtained by taking appropriate
derivatives. In this work we calculate the impurity entropy
$
S_\mathrm{imp}=S-S^{(0)},
$
where $S$ is the entropy of the full system, while $S^{(0)}$ is the
entropy of the conduction band alone. Using the NRG one can compute
$S_\mathrm{imp}$ over a wide range of temperature scales with little
numerical effort. To the contrary, the CTQMC becomes increasingly
numerically demanding at low temperatures and for larger hybridization
strength $\Gamma$. The comparison of the results can thus only be
performed in a limited temperature window which depends on $\Gamma$.
To circumvent this limitation, we perform calculations for a
range of $\Gamma$ at fixed $U$; in this way we tune the characteristic
low-energy scales of the problem (the Kondo temperature, the RKKY
scale) from very small to very large values, making them pass through
the available temperature window for different values of $\Gamma$.

In NRG, the impurity entropy is computed in the standard way
\cite{bulla2008}. To obtain good results even on the temperature scale
of the bandwidth, the discretization is performed with a small value
of the discretization parameter $\Lambda=1.8$ and the twist-averaging
with $N_z=16$ is used. Nevertheless, some small quantitative
systematic errors due to the discretization of the band are expected
for temperatures approaching the bandwidth. The truncation is
performed with a sufficiently high energy cutoff 
$E_\mathrm{cutoff}=12\omega_N$ that the results can be considered as
fully converged.

In CTQMC, we calculate the impurity energy for a range of
temperatures and obtain the entropy as 
\begin{equation}                                                                                                
S_\mathrm{imp}=\int \frac{1}{T} \left( \frac{\partial E_\mathrm{imp}}{\partial T} \right)_\mu
dT + \text{const}.                                                                                              
\end{equation}
The integration constant is fixed by the high-temperature asymptotic
limit of $2\ln4$. The impurity energy is defined as
$E_\mathrm{imp}=\langle H_\mathrm{dots} \rangle+E_\mathrm{c}$ with
\begin{equation}
E_\mathrm{c} = \expv{H_\mathrm{hyb}}+
\left( \expv{H_\mathrm{band}} - \expv{H_\mathrm{band}}_0 \right).
\end{equation}
The notation $\expv{}_0$ indicates that the expectation value is
computed for the system without the impurity (i.e., in the $\Gamma \to
0$ limit). $E_c$ is obtained from the impurity spectral function in
the Matsubara space $\GG(i\omega_n)$:
\begin{equation}
E_\mathrm{c}= \frac{\Gamma}{\beta} \sum_{i\omega_n,\sigma}                                                                      
d(i\omega_n)
\GG(i\omega_n), 
\end{equation}
where 
\begin{equation}
d(z)= \frac{-i\sgn[\Im(z)]}{\sqrt{1-z^2}}.   
\end{equation}
The impurity spectral function is defined as $\GG =
\GG_{11}+\GG_{12}+\GG_{21}+\GG_{22}$ with $\GG_{ij}(z)=\langle \langle
d_{i} ; d_{j}^\dag \rangle\rangle_z$. To avoid the minus sign problem,
the simulation is performed in the even/odd basis.
Since the asymptotic behavior of the Green's
function is $\GG(i\omega_n) \sim \frac{1}{i\omega_n}$, we subtract and
add $1/(i\omega_n)$
and calculate the problematic part exactly. 

\begin{figure}[htbp]
\centering
\includegraphics[clip,width=8cm]{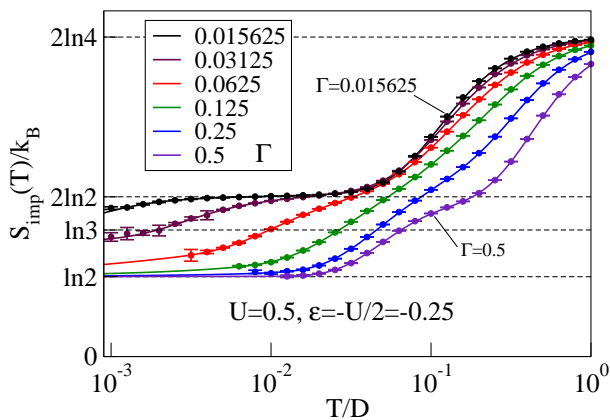}
\caption{(Color online) Comparison of the impurity entropy curves at
the particle-hole symmetric point. Lines: NRG, circles: QMC (with
error bars). Energy unit is the half-bandwidth $D=1$.} 
\label{fig1}
\end{figure}

\begin{figure}[htbp]
\centering
\includegraphics[clip,width=8cm]{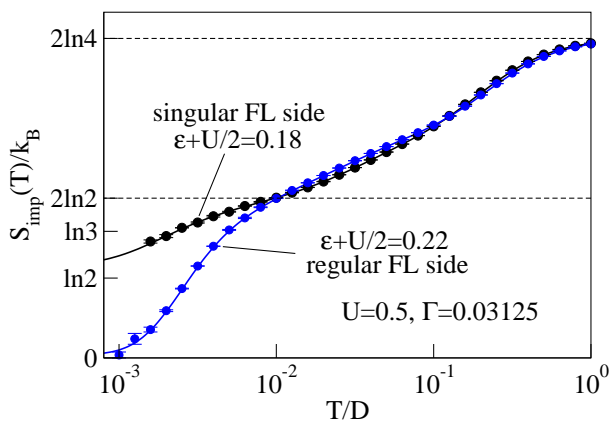}
\caption{(Color online) Comparison of the impurity entropy curves 
away from the particle-hole symmetric point. Lines: NRG, circles: QMC
(with error bars).
}
\label{fig1b}
\end{figure}

The results for the temperature dependence of the impurity entropy are
shown in Fig.~\ref{fig1} at the p-h symmetric point and in
Fig.~\ref{fig1b} away from it.  The agreement between the NRG and
CTQMC results is excellent in all cases considered.  
A small systematic deviation of the NRG from the QMC results at very
high temperatures ($T \sim D$) is anticipated. At intermediate
temperatures, the agreement improves until at some low
$\Gamma$-dependent temperature the QMC simulation can no longer be
performed in reasonable time due to slow thermalization. 
Nevertheless, NRG and QMC are found to agree below all the relevant
energy scales in the problem.
The DQD system near the p-h symmetric point thus behaves as a singular
FL, as predicted by the NRG. Furthermore, in Fig.~\ref{fig1b} we show
numerical evidence of a quantum phase transition (QPT) where, as a
function of $\epsilon$, the system goes from a singular FL to a
regular FL ground state (the BA solution predicts no such transition).

\begin{figure}[htbp]
\centering
\includegraphics[clip,width=8cm]{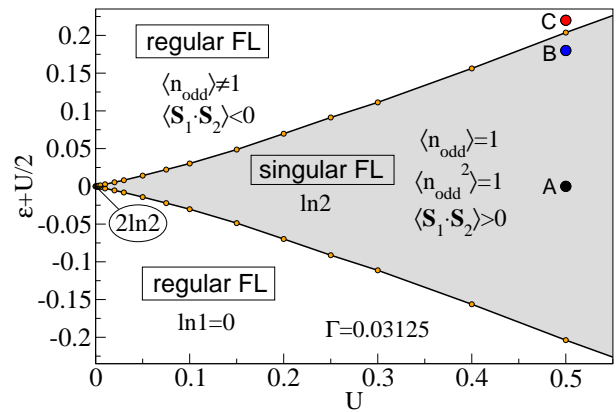}
\caption{(Color online) Phase diagram in the $(\epsilon,U)$ plane obtained
using the NRG. The entropy curve at point A is shown in Fig.~\ref{fig1},
while those at points B and C are compared in Fig.~\ref{fig1b}.}
\label{fig2}
\end{figure}

The DQD problem being integrable, this raises the question why the BA
solution differs from the NRG and QMC results. One possibility is that
the BA wave-function corresponds to some excited state instead of the
actual ground state of the system. We remark that the NRG and QMC are
both grand-canonical-ensemble calculations in the thermodynamic limit,
thus the occupancy of the dots is automatically correctly determined,
while in BA the thermodynamic limit is taken at the end of the
calculation, thus one needs to take care to choose the wavefunction
from the correct charge and spin sector. In particular, the occupancy
of the odd state $d^\dag_{o,\sigma}$ is important in this problem. In
the non-interacting limit, one has $[H,n_{\mathrm{odd}}]=0$ with
$n_{\mathrm{odd}}=\sum_\sigma d^\dag_{o,\sigma} d_{o,\sigma}$, thus
the occupancy of the odd state is a conserved quantity (the system is
non-ergodic). This state is completely decoupled from the continuum
and is sometimes referred to as the bound state in the continuum, dark
state, or ghost Fano resonance \cite{vzporedne2,guevara2003}. It lies
exactly at the Fermi level when the system is tuned to the
particle-hole symmetric point. Formally, at this point the system has
$2\ln 2$ residual entropy since for each spin the level may be either
occupied or unoccupied at no energy cost. As the interaction is turned
on, $n_{\mathrm{odd}}$ is no longer a constant of motion, yet the odd
state still plays a non-trivial role. We find that the ``dark state''
gives rise, at finite $U$, to a finite range of the gate voltages
$\epsilon$ around the particle-hole symmetric point where the
occupancy of the odd level is pinned exactly to one, i.e., the odd
level accommodates an electron of either spin, which is asymptotically
decoupled from the rest of the system on low energy scales. This gives
rise to a singular FL ground state with $\ln 2$ residual entropy.  The
breakdown of the regular FL ground state in this parameter range has
been noted previously using the Gunnarson-Sch\"onhammer variational
approach based on a regular FL trial function, which failed to
converge in the parameter range that is now known to correspond to the
singular FL phase \cite{tomi}. The corresponding interval of the gate
voltages $\epsilon$ is delimited by QPTs across which the impurity
charge changes discontinuously (in both even and odd levels)
\cite{vzporedne,vzporedne2}. These QPTs emerge directly from the
non-interacting dark state as the interaction is turned on, see
Fig.~\ref{fig2}.  We also note that the system is fully conducting at
the particle-hole symmetric point for any value of $U$, including
$U=0$. It remains an open technical problem how to include these
effects in the BA calculation in order to obtain the correct doublet
ground state around half filling.

In conclusion, we solved the problem of the parallel DQD using
hybridization expansion CTQMC and demonstrated that the results agree
with the results of the NRG. We pointed out the non-trivial role
played by the ``dark state'' odd orbital. We remark that the studied model
is an excellent benchmark test for various many-body techniques, since
only few appear to produce the correct results, and that even
supposedly exact analytical calculations need to be a-posteriori
validated against other reference results.

\begin{acknowledgments}
We acknowledge discussions with Sebastian Fuchs, Thomas Pruschke,
Toma\v{z} Rejec, Pascal Simon and David Logan. R. \v{Z}. acknowledges
the support of the Slovenian Research Agency (ARRS) under Grant No.
Z1-2058 and Program P1-0044.
\end{acknowledgments}

\bibliography{qmcnrg}

\end{document}